\documentclass[aps,prl,superscriptaddress,longbibliography,amssymb,twocolumn]{revtex4-1}
\usepackage[utf8]{inputenc}
\usepackage[pdftex]{graphicx}
\usepackage{amsmath}
\usepackage[amssymb,Gray]{SIunits}
\usepackage{physics}
\newcommand{\rmi}{\ensuremath{\mathrm{i}}}
\newcommand{\rme}{\ensuremath{\mathrm{e}}}
\newcommand{\rmd}{\ensuremath{\mathrm{d}}}

\newcommand{\nnl}{\nonumber\\}
\renewcommand{\vec}[1]{\ensuremath{\mathrm{\mathbf{#1}}}}
\newcommand{\R}{\ensuremath{\mathbb{R}}}
\newcommand{\C}{\ensuremath{\mathbb{C}}}
\newcommand{\Z}{\ensuremath{\mathbb{Z}}}

\begin{document}

\title{Nonlinear-ancilla aided quantum algorithm for nonlinear Schr\"odinger equations}


%
\author{Andr\'e Gro{\ss}ardt}
\email[]{andre.grossardt@uni-jena.de}
\affiliation{Institute for Theoretical Physics, Friedrich Schiller University Jena, Fr\"obelstieg 1, 07743 Jena, Germany}


\date{\today}

\begin{abstract}
We present an algorithm that uses a single ancilla qubit that can evolve nonlinearly, and show how to use it to efficiently solve generic nonlinear Schr\"odinger equations, including nonlocal Hartree equations and the Navier-Stokes equation for an irrotational, non-viscous flow. We propose a realization of such nonlinear qubits via spin-spin coupling of neutral atom qubits to a Bose-Einstein condensate. The results suggest that the use of nonlinear ancillas can provide substantial speedups compared to exclusively linear qubit devices.
\end{abstract}


\maketitle



The advent of fault-tolerant quantum computers promises considerable speedups for a variety of applications, most famously for factoring~\cite{shorPolynomialTimeAlgorithmsPrime1997} and searching~\cite{groverDifferentKindQuantum2005}. Linear systems can be efficiently solved~\cite{harrowQuantumAlgorithmSolving2009} with exponential speedup, which extends to the efficient solution of linear differential equations~\cite{caoQuantumAlgorithmCircuit2013,claderPreconditionedQuantumLinear2013,berryQuantumAlgorithmLinear2017,costaQuantumAlgorithmSimulating2019,engelQuantumAlgorithmVlasov2019,childsQuantumSpectralMethods2020,childsHighprecisionQuantumAlgorithms2021}. Even more straightforwardly, the linear Schrödinger equation can be simulated on a quantum computer~\cite{lloydUniversalQuantumSimulators1996,nielsenQuantumComputationQuantum2009} using only $\order{\log \mathcal{N}}$ qubits, $\mathcal{N}$ being the number of degrees of freedom or spatial grid points, as well as, a number of gates polynomial in both $\mathcal{N}$ and the number of time steps.

Due to the inherent linearity of quantum mechanical systems, efficient algorithms for nonlinear differential equations are considerably harder to find. Leyton and Osborne~\cite{leytonQuantumAlgorithmSolve2008} propose to use $k$ copies of a state to simulate $k$-th order polynomial functions in the probability amplitudes via postselection. This method can be used to solve systems of $\mathcal{N}$ nonlinear differential equations with polylogarithmic scaling in $\mathcal{N}$. It scales exponentially, however, with the number of time steps. Similar methods based on the reduction of effective qubits via postselection have been proposed as a means to introduce nonlinear operations~\cite{congQuantumConvolutionalNeural2019,beerTrainingDeepQuantum2020,hugginsVirtualDistillationQuantum2021,holmesNonlinearTransformationsQuantum2023}. Variational algorithms~\cite{lubaschVariationalQuantumAlgorithms2020} use related methods to approximate the ground state energy of nonlinear Schrödinger equations.
Other approaches rely on Koopman-von Neumann mapping of classical phase space dynamics to a unitary evolution of Hilbert space vectors~\cite{josephKoopmanvonNeumannApproach2020}, or employ Carleman linearization~\cite{liuEfficientQuantumAlgorithm2021}. The latter approach scales quadratically in time but the system size for the linearized equations scales with $\mathcal{N}^L$, where $L$ is the truncation level for the Carleman linearization determining the approximation error, and the method is only applicable for systems with strong dissipation.
Finally, the approach taken by Lloyd et al.~\cite{lloydQuantumAlgorithmNonlinear2020} is essentially reversing the mean field approximation for condensates, modeling nonlinear equations by considering the linear dynamics of a large number of identical states; it also scales quadratically in time but requires a qubit number that scales quadratically with the number of time steps as well. This requirement of a significantly larger Hilbert space than necessary for the simulation of corresponding linear equations poses a practical problem at least for noisy intermediate scale quantum devices.

Although quantum dynamics appears to be fundamentally unitary, nonunitary phenomena are not foreign to it in situations where large numbers of degrees of freedom are effectively reduced to a smaller number. The most prominent examples are open quantum systems, whose reduced dynamics are described by nonunitary master equations, and Bose-Einstein condensates (BECs), whose $N$-atom Hilbert space state can be approximated by a Hartree state $\ket{\Psi} = \bigotimes_{i=1}^N \ket{\psi}$, where the single atom wave function $\psi$ satisfies a nonlinear Schr\"odinger equation~\cite{grossStructureQuantizedVortex1961,pitaevskiiVortexLinesImperfect1961}.

The potential impact of nonlinearities on the capabilities of quantum computers has been understood early on~\cite{abramsNonlinearQuantumMechanics1998,czachorLocalModificationAbramsLloyd1998,czachorNotesNonlinearQuantum1998,childsOptimalStateDiscrimination2016}, when it was observed that a nonlinear evolution that drives quantum states apart exponentially fast would allow for polynomial time unstructured search that can be exploited to solve arbitrary NP problems in polynomial time. In this context, the applicability of the Gross-Pitaevskii nonlinearity~\cite{meyerNonlinearQuantumSearch2013} and generalizations thereof~\cite{meyerQuantumSearchGeneral2014} for unstructured search have been evaluated as well, finding that the required time resolution effectively results in resource requirements that still scale exponentially, even though providing a slight advantage over Grover's algorithm~\cite{groverDifferentKindQuantum2005}.

Here, we explore how the nonlinearities present in large quantum systems may be exploited for efficient solutions of nonlinear differential equations. We consider $n$-qubit states in the computational basis with $\mathcal{N} = 2^n-1$:
\begin{equation}
 \ket{\psi} = \sum_{k=0}^{\mathcal{N}} a_k \ket{k} \,,
\end{equation}
where $a_k = \braket{k}{\psi} \in \C$, and the binary representation of the integer $k$ corresponds to the computational basis state of an $n$-qubit system. We focus on the nonlinear Schr\"odinger evolution
\begin{equation}\label{eqn:nonlinear-schroedinger}
 \rmi \partial_t \ket{\psi}
= \hat{H}_\psi \ket{\psi} \,, \quad
\hat{H}_\psi = \hat{T}
+ \sum_{k,j = 0}^\mathcal{N} f_{kj} \abs{a_j}^2 \ketbra{k}{k} \,,
\end{equation}
with symmetric coefficients $f_{kj} = f_{jk} \in \R$, and $\hat{T}$ a self-adjoint operator for which we can efficiently simulate the unitary $\hat{U}_\epsilon = \rme^{-\rmi \epsilon \hat{T}}$ by a quantum circuit \footnote{Note that for $\hat{T} = f(\hat{p})$ a function of the momentum operator this can be achieved by a simple quantum Fourier transform of a diagonal unitary: $U_\epsilon = U_\text{QFT} \rme^{-\rmi \epsilon f(\hat{x})} U_\text{QFT}^\dagger$. In particular, this holds if $\hat{T} \sim \hat{p}^2$ is the kinetic energy operator.}.

Equation~\eqref{eqn:nonlinear-schroedinger} is equivalent to the nonlinear system
\begin{equation}
\rmi \frac{\rmd a_k}{\rmd t} = \sum_{j=0}^\mathcal{N} \left( T_{kj} a_j
+ f_{kj} \abs{a_j}^2 a_k \right)
\end{equation}
for the $2^n$ coefficient functions $a_k : \R \to \C$, where $T_{kj} = \bra{k} \hat{T} \ket{j} = T_{jk}^*$ are the matrix elements of $\hat{T}$ in the computational basis.

In addition to the $n$ principal qubits used to represent the state $\ket{\psi}$, for which we have a universal set of unitary gates, we assume the existence of a nonunitary ancilla qubit with the ability to apply both a controlled phase gate $CP(\lambda)$ and a controlled rotation $CR_x(\theta)$ with arbitrary angles $\lambda$, $\theta$ controlled on the $0$-th principal qubit. Using only $CR_x(\pi/n)$ and unitary gates on the principal qubits, we can construct~\cite{barencoElementaryGatesQuantum1995} an $n$-fold controlled-NOT gate out of $\order*{n^2}$ basic gates. For any $k \in \Z_{2^n}$, by applying $X$ gates to all qubits where $k$ has a 0-bit, followed by the $n$-fold controlled-NOT, we obtain a gate $CX_k$ that flips the ancilla as target qubit only for branches where the principal qubits are in the state $\ket{k}$.

In order to define the nonlinear capabilities of the ancilla qubits, note that any state of the $n+1$ qubit system can be written as
\begin{equation}\label{eqn:psi-sep}
 \ket{\Psi} = \alpha \ket{\psi_0} \ket{0}_a  + \beta  \ket{\psi_1} \ket{1}_a
\end{equation}
with $\abs{\alpha}^2 + \abs{\beta}^2 = 1$, where $\ket{0}_a$, $\ket{1}_a$ are the computational basis states for the ancilla qubit and $\ket{\psi_0}$, $\ket{\psi_1}$ are two arbitrary normalized states for the $n$ principal qubits. We then assume that there is a nonlinear single-qubit gate $N(\gamma)$ that maps the state \eqref{eqn:psi-sep} to
\begin{equation}\label{eqn:nonlin-trafo}
 \ket{\Psi} \mapsto \ket{\Psi'} = \rme^{\rmi \gamma \abs{\alpha}^2} \alpha \ket{\psi_0}\ket{0}_a  + \rme^{\rmi \gamma \abs{\beta}^2} \beta \ket{\psi_1}\ket{1}_a  \,,
\end{equation}
corresponding to a state dependent Bloch sphere rotation $(\theta, \varphi) \mapsto (\theta, \varphi - \gamma \cos \theta)$ around the $z$-axis.

In the following section we first introduce an algorithm that utilizes these nonlinear transformations in order to solve the nonlinear Schr\"odinger equation~\eqref{eqn:nonlinear-schroedinger}. We then propose a physical realization by coupling a neutral atom qubit to a BEC, followed by a discussion of applications for Hartree and Navier-Stokes equations.

\section{Algorithm}

We consider the initial state
\begin{equation}
 \ket{\Psi} = \sum_{k=0}^{\mathcal{N}} a_k \ket{k} \ket{0}_a
\end{equation}
in the computational basis $\bigcup_k \{\ket{k} \ket{0}_a, \ket{k} \ket{1}_a\}$. Acting with the $CX_k$ gate followed by the nonlinear gate and consecutive uncomputing, we then have
\begin{align}
\ket{\Psi} &\stackrel{CX_k}{\mapsto}
a_k \ket{k} \ket{1}_a
 + \sum_{j \neq k} a_j \ket{j} \ket{0}_a
 \nnl
&\stackrel{N(\gamma)}{\mapsto}
\rme^{\rmi \gamma \abs{a_k}^2} a_k \ket{k} \ket{1}_a
 + \rme^{\rmi \gamma (1 - \abs{a_k}^2)} \sum_{j \neq k} a_j \ket{j} \ket{0}_a
\nnl
&\stackrel{CX_k}{\mapsto}
\left( \rme^{2\rmi\gamma\abs{a_k}^2} a_k \ket{k}
 + \rme^{\rmi \gamma} \sum_{j \neq k} a_j \ket{j} \right) \ket{0}_a
\end{align}
where the last line is up to a global phase. Applying a controlled phase gate removes the phase $\rme^{\rmi \gamma}$.

Applying consecutively two such sequences for values $k$ and $l$ with nonlinear gates $N(\gamma_k)$ and $N(\gamma_l)$, respectively, followed by a sequence $CX_k \, CX_l \, N(\gamma_{kl}) \, CX_l \, CX_k$, we end up with coefficients
\begin{subequations}\begin{align}
a_k &\mapsto \rme^{2\rmi((\gamma_k + \gamma_{kl}) \abs{a_k}^2 + \gamma_{kl} \abs{a_l}^2)} a_k \\
a_l &\mapsto \rme^{2\rmi(\gamma_{kl} \abs{a_k}^2 + (\gamma_l + \gamma_{kl}) \abs{a_l}^2)} a_l \,.
\end{align}\end{subequations}
Iterating through all pairings $(k,l)$, we can then implement, with $\order*{\mathcal{N}^2 \log \mathcal{N}}$ gate complexity, the operator
\begin{equation}\label{eqn:wk-op}
 \hat{W}_\epsilon = \sum_{k=0}^\mathcal{N} \exp(-\rmi \epsilon \sum_j f_{kj} \abs{a_j}^2) \ketbra{k}{k} \,.
\end{equation}
for arbitrary symmetric coefficients $f_{kj} = f_{jk}$.

Trotterizing~\cite{lloydUniversalQuantumSimulators1996} equation~\eqref{eqn:nonlinear-schroedinger} as
\begin{equation}
 \ket{\psi_t} = \rme^{-\rmi t \hat{H}_\psi} \ket{\psi_0}
 \approx \left(\hat{U}_\epsilon \hat{W}_\epsilon \right)^{N_t} \ket{\psi_0} \,,
\end{equation}
with $N_t = \lfloor{t/\epsilon}\rfloor$, we obtain the final state at time $t$ from a given initial wave function $\psi_0$ up to an error of $\order*{\epsilon t}$.

The operator $\hat{U}_\epsilon$ is typically implemented with $\order*{\log \mathcal{N}}$ entangling gates~\footnote{For instance, if $\hat{T}$ is the kinetic energy operator, $U_\epsilon$ can be implemented via quantum Fourier transform with $\order*{n^2}$ gates}, resulting in a total gate complexity of $\order*{N_t \mathcal{N}^2 \log \mathcal{N}}$ entangling gates and $\order*{N_t \mathcal{N}^2}$ nonlinear gates. For a constant error, $N_t$ must scale quadratically with the time $t$. Contrary to previous algorithms~\cite{lloydQuantumAlgorithmNonlinear2020}, where the required qubit number also scales quadratically in $t$, this quadratic scaling can be achieved with a constant number of $n+1$ qubits.

Thus far, we only considered equation~\eqref{eqn:nonlinear-schroedinger} with a potential that is a linear function of the $\abs{a_i}^2$. Potentials that are higher order polynomials in $\abs{a_i}^2$ could be achieved by starting from multiple copies of the initial state. For instance, by applying our algorithm to the state $\ket{\psi} \otimes \ket{\psi} = \sum_{jk} a_j a_k \ket{j}\ket{k}$ terms proportional to $\abs{a_j}^2 \abs{a_k}^2$ are included in the state dependent phases of the operators~\eqref{eqn:wk-op}.

\section{Physical realization}

For a physical realization of the nonunitary qubit, consider a two component BEC~\cite{hallDynamicsComponentSeparation1998,byrnesMacroscopicQuantumComputation2012,byrnesMacroscopicQuantumInformation2015} with spin modes labeled by bosonic annihilation operators $\hat{a}_i$ and corresponding number operators $\hat{n}_i = \hat{a}^\dagger_i\hat{a}_i$ ($i=1,2$). Qubit states can be encoded via~\cite{byrnesMacroscopicQuantumComputation2012}
\begin{equation}
 \alpha \ket{0} + \beta \ket{1}
 = \frac{1}{\sqrt{N!}} \left(\alpha a_1^\dagger + \beta \hat{a}_2^\dagger\right)^N \ket{\text{vac}}
\end{equation}
in Fock space, which corresponds to first-quantized states $\ket{\Psi} = \ket{\varphi}^{\otimes N}$ with the single particle states~\cite{jiangParticlenumberconservingBogoliubovApproximation2016}
\begin{equation}
\ket{\varphi} = \frac{1}{N} \left(
\alpha \ket{\varphi_1} \otimes \ket{F_1} +
\beta \ket{\varphi_2} \otimes \ket{F_2} \right) \,,
\end{equation}
where $F_i$ corresponds to the two spin components.

The construction of controlled rotations from dipole-dipole interactions is well understood for ions and neutral atoms~\cite{beigeCoherentManipulationTwo2000, jakschFastQuantumGates2000} as well as between BECs~\cite{byrnesMacroscopicQuantumComputation2012} and can be straightforwardly extended to obtain the desired $CP(\lambda)$ and $CR_x(\theta)$ rotation gates for the nonunitary qubits controlled on the qubits of a conventional quantum computer, e.g.\ based on neutral atoms~\cite{barnesAssemblyCoherentControl2022,henrietQuantumComputingNeutral2020}.

We can implement the nonunitary gate~\eqref{eqn:nonlin-trafo} from the free evolution of the BEC. With the interaction Hamiltonian~\cite{wangSpinOrbitCoupledSpinor2010} $H_\text{int} = g_{11} \hat{n}_1^2 + g_{22} \hat{n}_2^2 + g_{12} \hat{n}_1 \hat{n}_2$, in units where $\hbar = m_\text{atom} = 1$, the spatial wave functions for the two modes then satisfy coupled equations~\cite{leeQuasispinModelMacroscopic2003,maimistovPotentialIncoherentAttraction1999,jiangParticlenumberconservingBogoliubovApproximation2016,liCoupledDensityspinBose2020}
\begin{equation}\label{eqn:bec-dynamics}
 \rmi \partial_t \varphi_i
 = \left(-\frac{1}{2} \partial_x^2 + V
 + \sum_j g_{ij} \abs{\varphi_j}^2
 \right) \varphi_i \,.
\end{equation}
If the system remains approximately in a stationary state for the trapping potential $V$, these dynamics induce a phase shift
\begin{subequations}\begin{align}
\alpha &\mapsto \rme^{\rmi t (g_{11} \abs{\alpha}^2
 + g_{12} \abs{\beta}^2)} \alpha
 = \rme^{\rmi t g_{12}} \rme^{\rmi t (g_{11} - g_{12}) \abs{\alpha}^2} \alpha \\
\beta &\mapsto \rme^{\rmi t (g_{12} \abs{\alpha}^2
 + g_{22} \abs{\beta}^2)} \beta
 = \rme^{\rmi t g_{12}} \rme^{\rmi t (g_{22} - g_{12}) \abs{\beta}^2} \beta \,.
\end{align}\end{subequations}
Hence, for a BEC with $g_{11} = g_{22}$ we can choose $t$ accordingly in order to find the desired map~\eqref{eqn:nonlin-trafo} up to an irrelevant global phase. For the case $g_{11} \ne g_{22}$ it is possible to adapt the algorithm for solving the nonlinear Schrödinger equation accordingly.

An alternative realization could be in optical quantum computers, if an appropriate coupling of photons to solitonic states~\cite{pareApproximateModelSoliton1990} can be implemented.

\section{Hartree and Navier-Stokes equations}

As a special case of equation~\eqref{eqn:nonlinear-schroedinger} we can consider the one-dimensional Hartree equations~\cite{frohlichMeanFieldLimitQuantum2004}
\begin{equation}\label{eqn:hartree}
 \rmi \partial_t \phi = -\partial_x^2\phi + V \phi \,,
 \quad V = \Phi \ast \abs{\phi}^2 \,,
\end{equation}
where $\phi : \R \times \R \to \C$ is a single particle wave function, $\Phi : \R \to \R$ a real-valued even function, and $\ast$ denotes the convolution of functions on $\R$. The Hartree equations~\eqref{eqn:hartree} constitute an entire class of nonlinear Schr\"odinger equations with applications in nonlinear fibre optics~\cite{agrawalNonlinearFiberOptics1995} or for the modelling of boson stars~\cite{ruffiniSystemsSelfGravitatingParticles1969}, among others. Choosing a spatial grid such that $\phi(t,x_0 + k \Delta x) = a_k$ we find
\begin{equation}
 V(x_0 + k \Delta x)
 = \sum_j f_{jk} \abs{a_j}^2
\end{equation}
with $f_{jk} = f_{kj} = \Phi((k-j)\Delta x) \Delta x$, and equation~\eqref{eqn:hartree} can be cast into the form~\eqref{eqn:nonlinear-schroedinger}. The extension to three dimensions is straightforward after an appropriate mapping of a spatial grid to integer values has been selected.

It may seem somewhat underwhelming to simulate a nonlinear Schr\"odinger equation using a physical system that itself obeys such an equation. It should be noted, however, that our method can be applied to the simulation of equation~\eqref{eqn:hartree} for \emph{arbitrary} functions $\Phi$ and \emph{arbitrary} initial states, regardless of the nature of the interaction and state of the BEC. In a BEC, $\Phi$ is generally assumed to be a point like interaction, $\Phi(\vec x) \sim \delta(\vec x)$. The Hartree equation then limits to the Gross-Pitaevskii equation~\cite{grossStructureQuantizedVortex1961,pitaevskiiVortexLinesImperfect1961}, whereas the methods presented are applicable to arbitraryly widespread nonlocal interactions $\Phi$.

As a second example, we can consider the Navier-Stokes momentum equation for vanishing viscosity and vorticity,
\begin{equation}\label{eqn:ns}
 \left(\partial_t + \vec u \cdot \nabla \right) \vec u
 = - \frac{\nabla p}{\rho} + \vec g \,.
\end{equation}
It has been noticed already by Madelung~\cite{madelungQuantentheorieHydrodynamischerForm1927} that this equation follows from the Schrödinger equation with the wave function ansatz $\phi = \sqrt{\rho} \exp(-\rmi \varphi)$ and the flow velocity $\vec u = \nabla \varphi$. However, in order to recover the right-hand side of equation~\eqref{eqn:ns} for a general space-time dependent pressure and density distribution the potential must obey the wave function dependent identity
\begin{equation}\label{eqn:ns-pot}
 \nabla V = \frac{\nabla p}{\rho} - \vec g
 + \nabla \frac{\nabla^2 \sqrt{\rho}}{2\sqrt{\rho}} \,,
\end{equation}
which results in a nonlinear dependence on $\rho$. In absence of the pressure gradient and external forces, $\nabla p = \vec g = 0$, we can expand around some static density $\rho_0$ and discretize the Laplace operator, obtaining
\begin{equation}\begin{split}
 V(t, \vec x) = \frac{1}{4 \rho_0 \, \Delta x^2} \sum_{i=1}^3
 \Big( \rho(t, \vec x + \Delta x \vec e_i) \\
 + \rho(t, \vec x - \Delta x \vec e_i)
 - 2 \rho(t, \vec x) \Big) + \order{\Delta x^2} \,,
\end{split}\end{equation}
where $\Delta x$ is the spatial grid size and $\vec e_i$ the unit vectors in $x_i$ direction. This is a nonlocal, linear function in $\rho$, to which our algorithm can be directly applied.

\section{Discussion}

We have shown that, for example, by coupling a single two-component BEC to atom qubits we can implement a nonlinear ancilla qubit, allowing for state dependent phase transformations. These phase transformations can in turn be used to simulate nonlinear Schr\"odinger equations with potentials that can be arbitrary symmetric polynomials of the square amplitudes $\abs{\braket{k}{\psi}}^2$. This is achieved with a considerably lower gate count than previously proposed algorithms for nonlinear problems on linear qubit devices.

This speedup is intuitively easily explained, because the true system size is not $n+1$ but rather $n+N$, where the number $N$ of atoms can easily surpass the number $n$ of principal qubits. Yet, the system is substantially easier to control than an $n+N$ qubit quantum computer would be. In this sense, the ancilla plays a role akin to an analog coprocessor.

As a word of caution, one must keep in mind that the nonlinear dynamics~\eqref{eqn:bec-dynamics} are only an approximation, which must be assured to hold in practical realizations, e.g., by choosing a sufficiently large number of atoms. Furthermore, the realization of the nonlinear gate proposed here will likely be slow compared to unitary gates, hence time complexity of the algorithm may be less favorable than the gate complexity suggests.

Be that as it may, for the area of noisy intermediate scale devices reduction of the required qubit resources can be of higher importance than runtimes for individual gates. The main purpose of this work is to emphasize the potential speedup one may gain from a nonlinear ancilla. In this regard, the results highlight such systems as an interesting option, in particular for solving nonlinear problems. An important question is, whether the algorithm presented here can be generalized to a broader class of problems. An interesting further course of study could be applications for quantum error correction.

\section*{Acknowledgments}
The author gratefully acknowledges funding through the Volkswagen Foundation.

\end{document}